\documentclass[twocolumn,showpacs,preprintnumbers,amsmath,amssymb]{revtex4}
\usepackage{dcolumn}

\usepackage[usenames,dvipsnames]{color}

\usepackage{graphicx}

\usepackage{amssymb}
\usepackage{amsmath}
\usepackage{bm}
\usepackage{epsfig}
\usepackage{ulem}

\newcommand {\be}{\begin{equation}}
\newcommand {\ee}{\end{equation}}
\newcommand {\bea}{\begin{eqnarray}}
\newcommand {\eea}{\end{eqnarray}}
\newcommand {\FIG}[1]{Fig. \ref{#1}}
\newcommand {\FIGS}[1]{Figs. \ref{#1}}
\newcommand {\EQ}[1]{Eq. (\ref{#1})}

\newcommand {\PRE}[1]{{Phys. Rev. E} {\bf {#1}}}

\newcommand {\PRL}[1]{{Phys. Rev. Lett.} {\bf {#1}}}

\newcommand {\SCI}[1]{{Science} {\bf {#1}}}

\newcommand {\JPCS}[1]{{J. Phys.: Conf. Ser.} {\bf {#1}}}

\begin{document}

\title{Explosive percolations on the Bethe Lattice}
\author{Huiseung Chae}
\author{Soon-Hyung Yook}\email{syook@khu.ac.kr}
\author{Yup Kim} \email[Corresponding author:]{ykim@khu.ac.kr  }
\affiliation{Department of Physics and Research Institute for Basic
Sciences, Kyung Hee University, Seoul 130-701, Korea}
\date{\today}

\begin{abstract}

Based on the self-consistent equations of the order parameter $P_\infty$ and the mean cluster size $S$, we develop a novel self-consistent simulation (SCS) method for arbitrary percolation on the Bethe lattice (infinite homogeneous Cayley tree). By applying SCS to the well-known percolation models, random bond percolation and bootstrap percolation, we obtain prototype functions for continuous and discontinuous phase transitions. By comparing the key functions obtained from SCSs for the Achlioptas processes (APs) with a product rule and a sum rule to the prototype functions, we show that the percolation transition of AP models on the Bethe lattice is continuous regardless of details of growth rules.

----

\end{abstract}
\maketitle


Since Achlioptas. et al. \cite{Ach} suggested an explosive percolation model,
there have been intensive studies on the explosive percolations (EP)
 \cite{daCosta10,Ziff09,Cho09,Radicchi,YKim10,Lee11,Grassberger11,Cho10,Woosik10,Fried,DSouza}.
Achlioptas process (AP) was originally argued to show the discontinuous phase transition
on the complete graph (CG) by suppressing growth of 
large clusters \cite{Ach}.
Subsequent studies on  variants
of EP models on networks and 
lattices also argued to show the discontinuous transition 
\cite{Cho09,Radicchi,Cho10,YKim10,Ziff09,Woosik10,Fried,DSouza}.
In contrast Riordan and Warnke \cite{Rio} analytically showed that the 
phase transition in AP model \cite{Ach} on CG is continuous
by use of the  arbitrary connectivity of CG. Furthermore several studies also showed
that the transition of variants of EP models on CG is continuous \cite{daCosta10, Grassberger11, Lee11}.
However, EP on CG was still reported
to undergo discontinuous transition
depending on the detail of  cluster growth rule \cite{Cho1st,D'Souza10}.
Therefore the transition nature of EP  on CG is not still
clear. Since the dimensionality of CG is infinite,
physics on CG must satisfy the mean-field theory. In this sense the mean-field theory of explosive percolation is not still clearly understood.

The Bethe lattice (infinite homogeneous Cayley tree) is physically a very important substrate or medium 
on which the mean-field theories for various physical models become exact \cite{Tho}. The analytic treatments of magnetic models \cite{orig},
percolation \cite{Stauffer_book,Tho} and localization \cite{Tho} on the Bethe lattice give important physical insights to the
subsequent developments of the corresponding research fields.
One of theoretical merits of the Bethe lattice is that one can setup the exact self-consistent equations (SCEs). In this letter, by use of the exact SCEs we develop a novel self-consistent simulation (SCS) method
for arbitrary percolation process on the Bethe lattice. 
From SCS method, we precisely calculate
the order parameter $P_{\infty}$ and the average size $S$ of finite clusters of
the AP models with a product rule or a sum rule.
The obtained $P_{\infty}$ and $S$ can clarify
the transition nature of AP models in the infinite dimension exactly. Furthermore unlike AP on CG, the bond connections on the Bethe lattice are purely local. 
Since there have been some papers that EP models on lattices with local bond connections
show the discontinuous transition \cite{Woosik10,Ziff09,Radicchi}, it is physically important to study EP models on the Bethe lattice or in the mean-field level with local connections.  
As we shall see, the transition of AP on the Bethe lattice is continuous regardless of the sum rule or the product rule. 
This result physically means that AP models with local connections in the mean-field level 
shows the continuous transition irrelevant to the details of growth rules.
In this sense the study on the Bethe lattice should give new physical insights to the mean-field theory of AP. 
\begin{figure}[ht]
\includegraphics{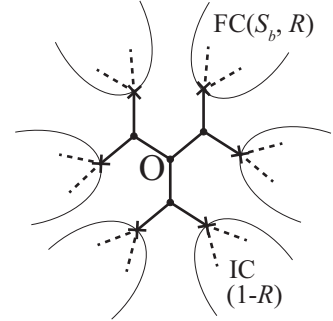}
\caption{Schematic diagram for a Bethe lattice with $z=3$.
The center part of the Bethe lattice is a 2-generation Cayley tree with 6 edge sites denoted by "X". Each edge site is connected to an infinite branch. 
Each infinite branch has an infinite cluster (IC) with the probability $1-R(=A)$. An infinite branch has a finite cluster (FC)
of the average size $S_b$ with the probability $R$.} \label{Bethe_fig}
\end{figure}

From the structure of the Bethe lattice, SCEs for macroscopically measurable physical quantity
can be easily setup \cite{Tho}. Let's first setup SCEs for arbitrary percolation on the Bethe lattice.
To setup SCE on the Bethe lattice in \FIG{Bethe_fig},
one starts with a center site (or origin {\bf O}) having $z$ bonds.
Then let's think about a part of the Bethe lattice with $n$ generations from {\bf O}, which have total
$N=1+z(k^n-1)/(k-1)$ sites with $k = z-1$. To make complete Bethe lattice, 
one should add an infinite
branch to each of $zk^{n-1}$ edge sites. To find the probability that {\bf O} 
belongs to the infinite cluster,
we need to know the occupation probability
$p$ of a bond (or a site) and the probability $R$ with which an edge site does not connected
to the infinity through the infinite branch
connected to it or the branch connected to the edge site has only finite clusters.
Then the order parameter $P_{\infty}$ of an arbitrary percolation,
which is defined by the probability of {\bf O} to
belong to the infinite cluster, is a function of $n$, $p$ and $R$ as $P_{\infty}(p,R,n)$.
Then SCE for $P_{\infty}$ is
\begin{equation}\label{selfc}
P_{\infty}=P_{\infty}(p,R,n) = P_{\infty}(p,R,n^\prime)
\end{equation}
for any combination of $\{n,n'\}$.
For the random or normal site percolation \EQ{selfc} with the combination
$\{n=1,n'=2\}$ gives $R=1-p +p R^k$ \cite{Stauffer_book,Tho},
which analytically reproduces the mean-field properties of percolation transition.
If the probability that a  cluster including {\bf O} with $s$ sites and $t$ edge sites within the
$n$-generation tree 
is $P(p,s,t,n,R,S_b)$, then
\begin{equation}\label{inft}
P_{\infty}(p,R,n) = 1 - \sum_{t} R^{t} \sum_{s} P(p,s,t,n,R,S_b),
\end{equation}
where $S_b$ is the average size of finite clusters connected to an edge site of the
$n$-generation tree in \FIG{Bethe_fig}. SCE for the average size $S$ of the finite clusters including {\bf O} can also be written as
\begin{equation}\label{mean_selfc}
S=S(p,R,S_b,n) = S(p,R,S_b,n').
\end{equation}
$S(p,R,S_b,n)$, where
\begin{equation}\label{mean}
S(p,R,S_b,n) = \frac{\sum_{s,t}P(p,s,t,n,R,S_b) \left[ s+tS_b \right] R^{t}}{1-P_{\infty}}.
\end{equation}

If one cannot calculate $P(p,s,t,n,R,S_b)$ analytically
or if one need to know $R$ and $S_b$ apriori to occupy a new bond (or site),
one should estimate $P(p,s,t,n,R,S_b)$ indirectly to  solve SCEs (\ref{selfc})
and (\ref{mean_selfc}). One of the indirect methods is a simulation method. In this letter we develop a simulation method to solve SCEs, 
which we call the self-consistent simulation (SCS).
In SCS, $P(p,s,t,n,R,S_b)$ is estimated by the relation $P(p,s,t,n,R,S_b) = {N(p,s,t,n,R,S_b)} / {N_{cluster}}$,
where $N(p,s,t,n,R,S_b)$ is the number of clusters including {\bf O} with $s$ sites and $t$ edge sites within the $n$-generation tree occurred in the simulation runs. Of course $N_{cluster}$ is the total number of clusters which includes {\bf O}
within the $n$-generation tree occurred in the same simulation runs. 
In the simulation both $P(p,s,t,n,R,S_b)$ and $P(p,s,t,n',R,S_b)$ are estimated simultaneously using the Bethe lattice with the center $n$-generation tree if $n > n'$. 
Since we don't know $R(p)$ and $S_b(p)$ apriori, the iteration processes are needed in SCS.
From initially guessed values for $R(p)$ and $S_b(p)$, the final or saturated values of $R(p)$ and $S_b(p)$ are obtained by the iteration of unit simulation process. The unit simulation process consists of the following two steps I) and II).
I) By use of the simulation runs based on the given values $R(p)$ and $S_b(p)$  $P(p,s,t,n,R,S_b)$ and $P(p,s,t,n',R,S_b)$ are
estimated. II) From the estimated $P(p,s,t,n,R,S_b)$, new $R(p)$ and $S_b(p)$ are calculated by utilizing SCEs (\ref{selfc}) and
(\ref{mean_selfc}).
In the unit simulation process to get the new $R(p)$ and $S_b(p)$, the quantities like $P(p,s,t,n,R,S_b)$ are estimated by
averaging over at least $10^6$ simulation runs. 
Such unit process is repeated until $R(p)$ and $S_b(p)$ reach the saturation values.
Using the saturated values of $R(p)$ and $S_b(p)$, $P_{\infty}$ and $S$ are estimated from 
Eqs. (\ref{inft}) and (\ref{mean}).
\begin{figure}[ht]
\includegraphics[width=8.5cm]{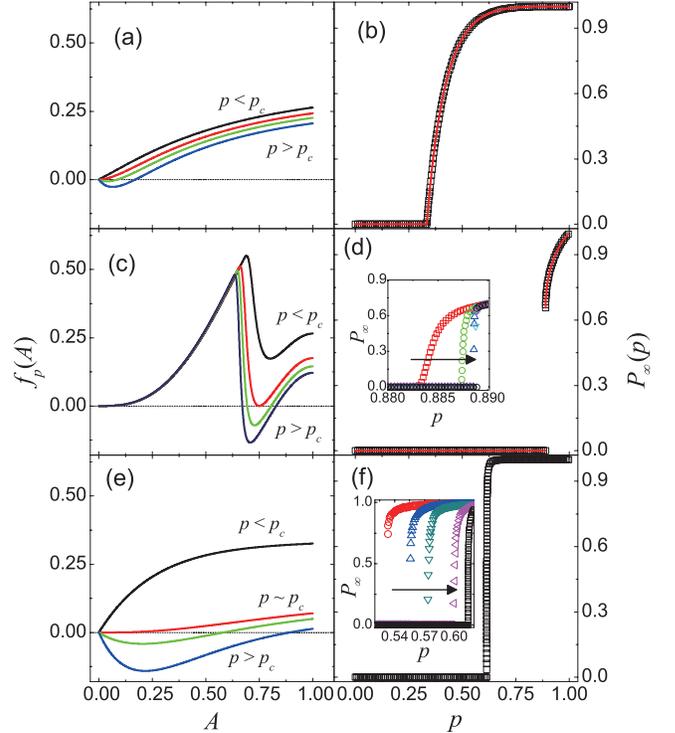}
\caption{(color online) SCS results of $f_p(A)$ and $P_\infty$ for random bond percolation (RBP),
bootstrap site percolation (BSP) and AP on the Bethe lattice with $z=4$. 
(a) $f_p(A)$ of RBP for $p=0.31333 (<p_c)$, $p=0.33333(\simeq p_c=1/3)$,
$p=0.34333(> p_c)$ and $p=0.35333(> p_c)$.
(b) Plot of $P_\infty$ for RBP against $p$. The line denote the analytic result. 
(c) $f_p(A)$ of BSP for $p=0.87888 (<p_c)$, $p=0.88888(\simeq p_c=8/9)$,
$p=0.89388(> p_c)$ and $p=0.89888(> p_c)$.
(d) Plot of $P_\infty$ for BSP against $p$. The line denote the analytic result. 
 Inset shows the evolution of results of iteration processes. The arrow $\rightarrow$ denotes the direction from the
earlier iteration process to the later iteration process.
(e) $f_p(A)$ of AP with a product rule (PR) for $p=0.60575 (<p_c)$,
$p=0.61575(\simeq p_c)$, $p=0.61775(> p_c)$ and $p=0.62575(> p_c)$.
(f) Plot of $P_\infty$ for AP with PR against $p$. 
   Inset shows the evolution of results of iteration processes.  The arrow $\rightarrow$ means the same thing as in (d).
} \label{PRG}
\end{figure}

The phase transition of random or normal percolation (RBP) on the Bethe lattice
is well-known to be continuous \cite{Stauffer_book}. 
We have applied our SCS to random bond percolation on the Bethe lattice with
$z=4$.
The results for $z=4$, $n=13$, $n'=5$ are displayed in \FIGS{PRG}(a) and (b).
In \FIG{PRG}(a) we display the simulation results for

\bea
 f_p(A)\equiv P_{\infty}(p,1-A,n') - P_{\infty}(p,1-A,n),
 \eea
where $A \equiv 1-R$ for various $p$. From \EQ{selfc}, $A^*$ (or $R^*$)
 which satisfies $f_p(A^*)=0$ is the real physical value for a given occupation probability $p$.
 For $p<p_c$ there occurs only trivial
 solution $A^*=0$ as shown in \FIG{PRG}(a).
 Increasing $p$ from $p_c$, the nontrivial solution $A^*>0$
 continuously increases from zero.
 This continuous increase  makes the order parameter $P_\infty(p)$ increase continuously as in \FIG{PRG}(b).  
The simulation result for $P_\infty(p)$ exactly coincides
with the analytic result
$P_{\infty} = 1 - [1/2 - \sqrt{(4-3p)/4p} ]^4$ for $p>p_c (=1/k=1/3)$ as show in \FIG{PRG}(b). 
Therefore $f_p(A)$ which behaves like in \FIG{PRG}(a) is a prototype
function for the continuous transition. 

Bootstrap site percolation (BSP) on the Bethe lattice \cite{Boot1} is analytically known to show the discontinuous transition. In BSP
the order parameter $P_\infty$ is the probability for an occupied site to be a site of the infinite $m$-cluster.
 Here the $m$-cluster means the cluster in which every occupied site has at least $m$
 occupied nearest neighbors.
 For $m \ge 3$ the phase transition of BSP on the Bethe lattice is discontinuous \cite{Boot1}.
 By using SCS we have obtained $f_p(A)$ and $P_\infty$
 for BSP with $m=3$ on the Bethe lattice with $z=4$.
The results of SCS for BSP with $z=4$, $m=3$, $n=13$, $n'=1$
are depicted in \FIGS{PRG}(c) and (d).
As shown in \FIG{PRG}(c), $f_p(A)=0$ for $p<p_c$ has only the trivial solution
$A^*=0$ as in RBP. 
In contrast to RBP the nontrivial solution of $f_p(A)=0$ for $p>p_{c}$ comes from the peculiar
behavior of $f_p(A)$, which reminds us the thermodynamic instability
in the thermal mean-field first order transition, which makes the sudden
jump of the $A^*$ from zero at $p=p_c=8/9$.
The jump of $A^*$ causes the discontinuous increase of
$P_\infty$ as shown in Fig. 2(d), which is exactly the same as the analytic result, 
$P_{\infty} = p(1-R)^{4} + 4pR(1-R)^{3}$ with $R=1$ for $p<p_c$ and $R=[1 - 3 \sqrt{1-p_c/p}]/4$ for $p>p_c$ \cite{Boot1}.
Therefore $f_p(A)$ like in Fig. 2(c) is a prototype function for the discontinuous transition. 
Furthermore, as shown in the inset of Fig. 2(d), the iteration processes in SCS for
the BSP with the initial value $R=0$ drives $P_\infty(p)$
from the continuous increase to the final discontinuous jump at $p=p_c$, which is also a typical behavior of the discontinuous transition
on the Bethe lattice.

We now focus the AP model. To occupy a bond in the AP model \cite{Ach},
two bonds $\alpha$ and $\beta$ are randomly chosen.
Let $S_{\alpha 1}$ and $S_{\alpha 2} (S_{\beta 1}$ and $S_{\beta 2})$
be the sizes of the two clusters which would be connected by occupying bond $\alpha$ ($\beta$).
Under a product rule (PR), the bond $\alpha$ is chosen for the growth of clusters and $\beta$ is discarded
if $\prod_{j=1}^{2} S_{\alpha j} < \prod_{j=1}^{2} S_{\beta j}$.
Otherwise, the bond $\beta$ is chosen. 
AP model with a sum rule (SR) is the same as that with PR
except for the change of the condition 
into $\sum_{j=1}^{2} S_{\alpha j} < \sum_{j=1}^{2} S_{\beta j}$.
Note that arbitrary edge site is connected to an infinite cluster with the probability $A$ depicted as in \FIG{Bethe_fig}. Therefore $S_{\alpha(\beta)j}$ is calculated  as 
\begin{equation}\label{realsize}
S_{\alpha(\beta)j} = s_{\alpha(\beta)j} +\infty \times I_{\alpha ( \beta )j} + ( t_{\alpha(\beta)j} - I_{\alpha(\beta)j})S_{b} ~~, 
\end{equation}
where $s_{\alpha (\beta)j}$ is the number of sites within the $n$-generation tree in the cluster $\alpha(\beta)j$,
$t_{\alpha (\beta)j}$ is the number of edge sites 
and $I_{\alpha (\beta)j}$ is the number of edge sites which are connected to the infinite cluster.
Of course  $I_{\alpha (\beta)j}$ depends on $R$ or $A$.
Therefore $P(p,s,t,n,R,S_b) (= {N(p,s,t,n,R,S_b)} / {N_{cluster}}$)
depends apriori on $R$ and $S_b$ and thus the iteration is essential in SCS for AP model.
For SCS of AP model,
it should be careful to choose $n'(<n)$ for a given $n$, because 
too small $n'$ cannot convey enough information for AP and $n'$ close to $n$
can never gives physically plausible solutions for SCEs (\ref{selfc}) and (\ref{mean_selfc}).
From the simulations with various $\{n,n^\prime\}$ it is confirmed that the suitable
choice of $n'$ should be in the interval $n/3 <n'<n/2$.  
$f_p(A)$ and $P_\infty$ for AP model with PR
obtained from SCS with $z=4$, $n=14$, $n'=5$ are displayed In \FIGS{PRG}(e) and (f).
 The results for AP model with SR are nearly the same as 
those in \FIGS{PRG}(e) and (f) except that $p_c$ for SR is slightly smaller than $p_c$ for PR.
As can be seen from  \FIG{PRG}(e), $f_p(A)=0$ has only trivial solution
$A^*=0$
for $p<p_c$ as RBP and BSP.
Increasing $p$ from $p_c$, the nontrivial solution
$A^* (> 0)$
continuously increase from zero.
$A^*$ for AP increases very rapidly compared to $A^*$ for RBP as $p$ increases.
Except for this rapid increase, $f_p(A)$ for AP behaves in 
the same way as that for RBP in \FIG{PRG}(a), which is 
the prototype function for the continuous phase transition.
We cannot find any symptom that $f_p(A)$ for AP behaves like $f_p(A)$ for BSP in Fig. 2(c). 
The continuous increase of $A^*$ makes the order parameter $P_\infty(p)$
for AP increase continuously as in \FIG{PRG}(f).
Moreover, as can be seen from the inset of \FIG{PRG}(f) the iteration processes in SCS for
AP with PR drives
$P_\infty(p)$ from the discontinuous jump to the final continuous increase at $p=p_c$,
contrary to those in the inset of \FIG{PRG}(d).

\begin{figure}[ht]
\includegraphics[width=8.5cm]{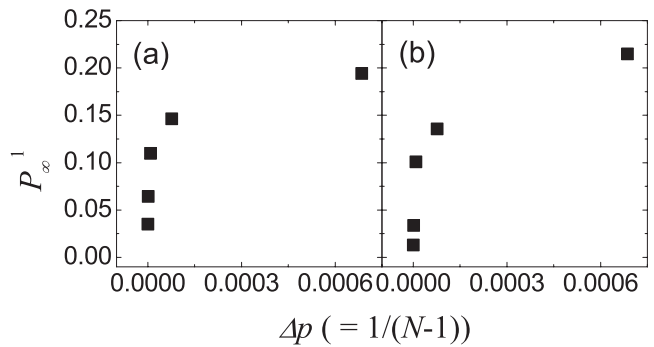}
\caption{${P_{\infty}}^{1}$ against $\Delta p(\equiv 1 / (N-1))$ for 
$\{n,n'\}=\{7,4\},\{9,4\},\{11,5\},\{13,5\},\{14,5\}$.
(a) AP model with PR and (b) the model with SR (b).
} \label{Gap}
\end{figure}
In SCS on the Bethe lattice with the $n$-generation center tree, we cannot obtain $P_\infty$ against all the continuous $p \in [0,1]$. Instead in SCS $p$ is increased
discretely by $\Delta p = 1/(N-1)$, where $N$ is the total number of sites
in the $n$-generation Cayley tree. Therefore in SCS the lowest nonzero  $P_\infty^{\phantom{0}1}$
occurs at $p=p'_c$ very close to the true $p_c$ with $0< (p'_c -p_c) \le \Delta p$.
In \FIG{Gap} we display the dependence of ${P_\infty}^1$ on
$\Delta p$ for SCSs of AP models with $\{n,n'\}=\{7,4\},\{9,4\},\{11,5\},\{13,5\},\{14,5\}$.
As can be seen from \FIG{Gap}, ${P_\infty}^1$ for both AP models with PR and SR
decreases monotonically to zero as $\Delta p$ decreases to zero.
This result for $P_{\infty}^{\phantom{0}1}$ also supports the fact
that the phase transition nature of AP on the Bethe lattice is continuous.

\begin{figure}[ht]
\includegraphics[width=9.0cm]{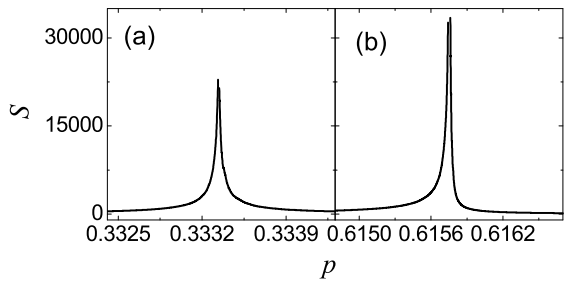}
\caption{$S$ on the Bethe lattice with $z=4$.
(a) Plot of $S$ for RBP against $p$. The data are obtained from SCS with $n=13$, $n'=5$. 
(b) Plot of $S$ for AP with PR against $p$. The data are obtained from SCS with $n=14$, $n'=5$.
} \label{SG}
\end{figure}

From SCS based on \EQ{mean_selfc}, the average size $S$ of the finite clusters is obtained for  
RBP and AP model with PR as in \FIG{SG}. $S$ for AP model with SR shows nearly the same behavior as \FIG{SG}(b).
It is confirmed that $S$ for RBP from SCS
is nearly identical to the analytic result
$S=1+4pR/(1-p-2pR)$, where $R=1$ for $p<p_c$ and $R = 1+\sqrt{(4-3p)/4p^3}-3/2p$ for $p>p_c$.
Even though $S$ for AP model diverges more rapidly than $S$ for RBP, $S$ for both RBP and AP diverges as 
$S(p) \simeq | p-p_c |^{-\gamma}$ with the susceptibility exponent
$\gamma=1.00(1)$. The result for $S$ of AP models also supports
that the transition in AP models is continuous
on the Bethe lattice.

Since the transition in the AP models is continuous, the order parameter exponent $\beta$ is calculated by fitting
the relation $P_\infty \simeq (p-p_c)^\beta$ to the data for $p>p_c$ but very close to $p_c$ in \FIG{PRG}(f). Since $f_p(A)$ for $p \simeq p_c$ in \FIG{PRG}(e) increases very slowly as $A$ increases from zero, $P_\infty$ increases very rapidly and the exponent $\beta$ is expected to be very small. From the best fit
the obtained exponent is $\beta = 0.05(5)$. The data even fits very well to $P_\infty  \simeq |\ln (p-p_c)|^{-\chi}$ with $\chi=3.4(1)$.
This result means that the exponent $\beta$ on the Bethe lattice is very small and nearly identical to those obtained
on the complete graph \cite{Grassberger11,Cho10,Radicchi}.

In summary, we show that AP models on the Bethe lattice have a continuous transition
from the SCSs developed to cover arbitrary percolation processes on the Bethe lattice. 
For this $f_p(A)$ for AP is first shown to be physically identical to that for RBP.
We also shown that ${P_\infty}^1$ for AP model decrease to zero as $\Delta p=1/(N-1)$ goes to zero.
The divergent behavior of $S$ for AP models is also shown to be the same as that for RBP with $\gamma=1$.
The exponent $\beta$ is also shown to be very small or $\beta \simeq 0.05$.

This work was supported by National Research Foundation
of Korea (NRF) Grant funded by the Korean Government
(MEST) (Grant Nos. 2009-0073939 and 2011-0015257).

\end{document}